\documentclass[english,sort&compress]{IEEEtran}
\usepackage[T1]{fontenc}
\usepackage[latin9]{inputenc}
\usepackage{array}
\usepackage{float}
\usepackage{multirow}
\usepackage{amsmath}
\usepackage{amssymb}
\usepackage{graphicx}
\usepackage{tikz}
\usepackage{lipsum}

\newcommand\copyrighttext{%
  \footnotesize © 2013 IEEE. Personal use of this material is permitted. Permission from IEEE must be obtained for all other uses, in any current or future media, including reprinting/republishing this material for advertising or promotional purposes, creating new collective works, for resale or redistribution to servers or lists, or reuse of any copyrighted component of this work in other works.}
\newcommand\copyrightnotice{%
\begin{tikzpicture}[remember picture,overlay]
\node[anchor=south,yshift=10pt] at (current page.south) {\fbox{\parbox{\dimexpr\textwidth-\fboxsep-\fboxrule\relax}{\copyrighttext}}};
\end{tikzpicture}%
}

\makeatletter

\providecommand{\tabularnewline}{\\}
\floatstyle{ruled}
\newfloat{algorithm}{tbp}{loa}
\providecommand{\algorithmname}{Algorithm}
\floatname{algorithm}{\protect\algorithmname}

\usepackage{dsfont}

\makeatother

\usepackage{babel}
\begin{document}

\title{Dispersive Media Subcell Averaging in the FDTD Method using Corrective
Surface Currents}

\author{Joachim Hamm, Fabian Renn, and Ortwin Hess}
\maketitle
\begin{abstract}
We present a corrective subcell averaging technique that improves
on the accuracy of the volume-averaged finite-difference time-domain
(FDTD) method in the presence of dispersive material interfaces\emph{.}
The method is based on an alternative effective-medium formulation
that captures field discontinuities at interfaces as electric and
magnetic surface currents. In calculating the spectra of strongly
dispersive Mie scatterers we demonstrate that the derived FDTD algorithm
is both highly efficient and able to approximately restore second
order accuracy. 
\end{abstract}

\section{Introduction}

Half a century after its invention by Kane Yee \cite{Yee1966} the
finite-difference time-domain (FDTD) method remains a popular choice
for simulating the propagation of electromagnetic waves and their
interaction with electronic media \cite{Yee1966,Taflove2005}. The
simplicity of the algorithm and its low computational footprint are
contrasted by the use of non-conformal grids, which, if field discontinuities
are not properly accounted for, reduce accuracy from second to first
order \cite{Hwang2001,Taflove2005}. This not only negates the advantage
of the staggered grid Yee-algorithm but also impacts on the computational
cost when modeling systems that exhibit geometric features on sub-wavelength
scales due to poor convergence. 

The problem of restoring accuracy of the FDTD scheme in the presence
of interfaces was first studied in the microwave regime \cite{Yu2001,Marcysiak1994,Kaneda1997}.
Since then a variety of effective-permittivity (EP) models have been
suggested for the treatment of field discontinuities at material interfaces,
which can broadly be classified as either contour-path (CP) or volume-polarized
(VP) models \cite{Fujii2003,Hirono2000,Mohammadi2005,Yu2001,Dey1999,Railton1999}.
Fundamentally, defining the effective permittivity $\tilde{\varepsilon}_{\infty}$
as volume-average (VA) of the permittivity $\varepsilon_{\infty}$
over one Yee-cell $\tilde{\varepsilon}_{\infty}=\langle\varepsilon_{\infty}\rangle$
is compatible with the standard FDTD scheme but does not constitute
an accurate VP model as discontinuities of the electric field at interfaces
are not accounted for. In this context the VP model proposed by Farjadpour
et al. \cite{Farjadpour2006} is of particular importance. Based on
the continuity of the parallel electric and normal displacement field
components, the effective permittivity tensor is derived as $\tilde{\boldsymbol{\boldsymbol{\varepsilon}}}_{\infty}^{-1}=\langle\varepsilon_{\infty}^{-1}\rangle\mathbb{P}+\langle\varepsilon_{\infty}\rangle^{-1}(\mathds1-\mathbb{P})$,
where $\mathbb{P}=\mathbf{n}\otimes\mathbf{n}$ performs a vector-projection
onto the face-normal of the interface. The application of this non-diagonal
and anisotropic permittivity tensor requires interpolation of the
Yee-centered $\mathbf{D}$-field to the cell-center and subsequent
interpolation of the cell-centered $\mathbf{E}$-field back onto the
Yee-grid, a procedure that effectively equates to a smoothing operation
with extended spatial stencil \cite{Werner2007}. Nonetheless, as
numerical evidence suggests, the spectral accuracy increases to approximately
second order, reducing the computational cost for problems that involve
non-dispersive dielectrics (e.g., photonic crystal applications).
In 2007, Deinega et al. \cite{Deinega2007} suggested an approach
that extends this method to the linear dispersive regime. Their algorithm
uses the decomposition $\mathbf{E}=\mathbf{E}_{||}+\mathbf{n}(E_{\perp,1}+E_{\perp,2})$,
where $E_{\perp}=\mathbf{n}\cdot\mathbf{E}$ and $\mathbf{E}_{||}=\mathbf{E}-\mathbf{n}(\mathbf{n}\cdot\mathbf{E})$,
to split the electric field into four independent components, which
drive the polarization currents at the interface. While the split-field
approach applies to the general case it is noteworthy that splitting
the electric field into normal and parallel components is not always
necessary. For example, Lee et al. \cite{Lee2010} derive a model
that uses an effective conductivity tensor in the quasi-static limit
without splitting the fields, while Liu et al. \cite{Liu2012} employ
a rotation of the coordinate system in conjunction with modified material
responses to avoid an explicit computation of the four split-field
components. The resulting algorithms are computationally more efficient
yet less general in the sense that they do not apply to arbitrary
dispersive material responses. Furthermore, as in \cite{Deinega2007},
it remains unspecified how these algorithms interface with the standard
Yee-centered algorithm that could be efficiently employed across regions
where permittivities are smooth. 

\begin{figure}
\begin{centering}
\includegraphics[scale=0.95]{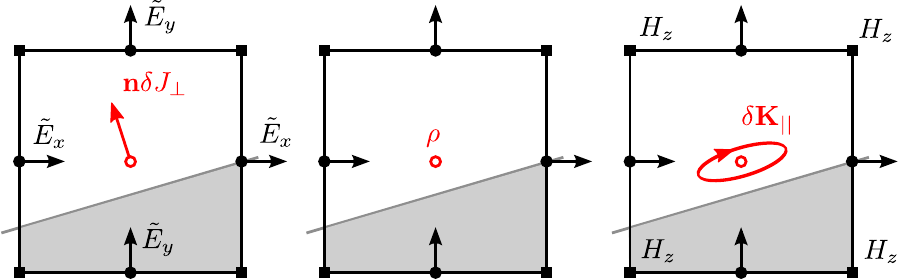}
\par\end{centering}

\caption{\label{fig:1}(color online) Representation of the VA+CC FDTD algorithm
in two dimensions. $\tilde{\mathbf{E}}$ and $\mathbf{H}$ fields
(black) are calculated using the standard VA FDTD algorithm, introducing
systematic errors at interface-cells due to discontinuities of the
field. The surface currents $\delta\mathbf{J}_{\perp}$ and $\delta\mathbf{K}_{||}$
(red) correct the errors during the electric (left) and magnetic (right)
update steps. Calculating the corrective currents requires an intermediate
step for integrating the surface charge density $\rho$ at the cell
center (middle). Note, that $\tilde{\mathbf{E}}$ is an approximate
field from which we can reconstruct the discretized electric field
$\mathbf{E}$. The algorithm is valid for any number of dimensions
(one to three) and compatible with the standard Yee scheme as the
dispersive corrections only apply to interface cells.}
\end{figure}

\copyrightnotice
Here, we present an alternative VP approach (see Fig. \ref{fig:1})
that solves the EP curl equations on the Yee-grid using the standard
volume-averaged FDTD algorithm but replaces the electric field with
an approximate field $\tilde{\mathbf{E}}$ that is continuous across
non-dispersive interfaces. The field discontinuities at dispersive
media interfaces need then to be captured as corrective electric and
magnetic currents $\delta\mathbf{J}_{\perp}$ and $\delta\mathbf{K}_{||}$,
which are induced by a surface charge field $\rho$. Based on this
idea we first formulate an effective medium theory and then show how
this EP model translates into a FDTD scheme that offers some unique
advantages: 1) the algorithm naturally extends the standard FDTD scheme
by introducing additive current corrections that only apply at interface
cells; 2) only the normal field components are subjected to spatial
smoothing operations at the interface; and 3) calculating the corrections
is computationally efficient and requires no alteration of the dispersive
response functionals (as for example in \cite{Liu2012,Lee2010}).
In the result section, we apply the derived algorithm to the example
of a highly dispersive Mie scatterer in two dimensions, demonstrating
stability and allowing for a comparison of numerical errors between
the VA+CC (using current corrections), the VA, and standard staircasing
schemes.

\section{Corrective-Current Subcell Smoothing}

Our starting point are the split-field equations derived by Deinega
et al. \cite{Deinega2007} (equations (3)-(6) therein). Without loss
of generality we write the scalar permittivity as $\varepsilon(\omega)=\varepsilon_{\infty}+\chi(\omega)$
and transform the equations into time-domain. Using a slightly different
notation, we write
\begin{equation}
\begin{aligned}\langle\varepsilon_{\infty}\rangle\partial_{t}\mathbf{E}_{||} & =\left(\nabla\times\mathbf{H}\right)_{\parallel}-f_{1}\mathbf{J}_{1}[\mathbf{E}_{||}]-f_{2}\mathbf{J}_{2}[\mathbf{E}_{||}]\\
\varepsilon_{\infty,1}\partial_{t}E_{\perp,1} & =f_{1}\left(\nabla\times\mathbf{H}\right)_{\bot}-J_{1}[E_{\perp,1}]\\
\varepsilon_{\infty,2}\partial_{t}E_{\perp,2} & =f_{2}\left(\nabla\times\mathbf{H}\right)_{\bot}-J_{2}[E_{\perp,2}]
\end{aligned}
\label{eq:DEINEGA}
\end{equation}

where $\mathbf{J}_{1/2}[\mathbf{E}]=\partial_{t}\mathbf{P}_{1/2}[\mathbf{E}]$
are functionals of $\mathbf{E}$, describing the (isotropic) polarization
current response. The symbols '$\perp$' and '$\parallel$' denote
vector-projections relative to the interface with face-normal $\mathbf{n}$
and the notation $J_{1/2}=\mathbf{n}\cdot\mathbf{J}_{1/2}$ is introduced
for brevity where quantities with a $\bot$ suffix are always scalars
(for example $E_{\bot}=\mathbf{n}\cdot\mathbf{E}$) and quantities
with a $\parallel$ are always vectors (for example: $\mathbf{E}_{\parallel}=\mathbf{E}-\left(\mathbf{E}\cdot\mathbf{n}\right)\mathbf{n}$).
In adopting vector-notation we do not impose restrictions on the numbers
of dimensions (i.e., the equations are valid for the two- and three-dimensional
case). The above formulation of Ampre's law implicitly assumes an
averaging over a volume-cell that is intersected by a boundary between
media $1$ and $2$ with cell-filling ratios $f_{1}$ and $f_{2}$
($f_{1}+f_{2}=1$). Angled brackets are used throughout this work
to denote volume averages of the form $\langle\varepsilon_{\infty}\rangle=f_{1}\varepsilon_{\infty,1}+f_{2}\varepsilon_{\infty,2}$.

The derivation of (\ref{eq:DEINEGA}) is straightforward but their
translation into an efficient and stable finite-difference scheme
is not. To retain second order accuracy, the field components in the
curl expression $\nabla\times\mathbf{H}$ should be calculated on
the Yee-grid while the projections onto parallel and normal projections
require interpolation to the cell-center. After calculating the updates
of the $\mathbf{E}_{||}$, $E_{\perp,1}$ and $E_{\perp,2}$ components
at the cell-center the $\mathbf{E}$-field thus needs to be reconstructed
and redistributed onto the Yee-grid. However, a direct implementation
proves impractical for the following reason. The cell-centered four-field
representation and the extended spatial stencil (due to interpolation
between the grids) is incompatible with the standard Yee-algorithm.
As a consequence the algorithm is best deployed across the whole grid
irrespective of whether cells are intersected by media-boundaries
or not. This introduces unnecessary smoothing operations across the
whole grid, increases the computational cost and requires a reimplementation
of the infrastructure typically associated with FDTD frameworks (e.g.,
total-field scattered-field injection, boundary conditions etc). 

We here seek to derive an alternative formulation where the standard
Yee scheme can be efficiently applied across the domain augmented
by corrections that only apply to the comparably small number of interface
cells. The basis for this corrective method is a reformulation of
(\ref{eq:DEINEGA}). In introducing new variables for the normal electric
field and the density of the induced surface charges, 
\begin{equation}
\begin{alignedat}{1}E_{\perp} & =E_{\perp,1}+E_{\perp,2}\\
\rho & =f_{2}\varepsilon_{\infty,1}E_{\perp,1}-f_{1}\varepsilon_{\infty,2}E_{\perp,2}
\end{alignedat}
\end{equation}
equations (\ref{eq:DEINEGA}) can be cast into the form
\begin{equation}
\begin{aligned}\langle\varepsilon_{\infty}\rangle\partial_{t}\mathbf{E}_{||} & =\left(\nabla\times\mathbf{H}\right)_{\parallel}-f_{1}\mathbf{J}_{1}[\mathbf{E}_{||}]-f_{2}\mathbf{J}_{2}[\mathbf{E}_{||}]\\
\langle\varepsilon_{\infty}^{-1}\rangle^{-1}\partial_{t}E_{\perp} & =\left(\nabla\times\mathbf{H}\right)_{\bot}-\zeta_{1}J_{1}[E_{\perp,1}]-\zeta_{2}J_{2}[E_{\perp,2}]\\
\partial_{t}\rho & =f_{1}J_{2}[E_{\perp,2}]-f_{2}J_{1}[E_{\perp,1}]
\end{aligned}
\label{eq:DEINEGA-1}
\end{equation}
with $\zeta_{1/2}=\langle\varepsilon_{\infty}^{-1}\rangle^{-1}\varepsilon_{\infty,1/2}^{-1}$.
The fact that $\langle\varepsilon_{\infty}\rangle^{-1}\ne\langle\varepsilon_{\infty}^{-1}\rangle$
makes it impossible to reconstruct Ampre's law in isotropic form
by directly combining the first two equations. However, we can define
an approximate electric field
\begin{equation}
\tilde{\mathbf{E}}=\mathbf{E}_{||}+\langle\varepsilon_{\infty}\rangle^{-1}\langle\varepsilon_{\infty}^{-1}\rangle^{-1}\mathbf{n}E_{\perp}
\end{equation}
which, in the absence of dispersive currents, is continuous across
material interfaces and matches $\mathbf{E}$ at non-interface cells.
Combining the first two equations of (\ref{eq:DEINEGA-1}) in this
fashion yields
\begin{equation}
\langle\varepsilon_{\infty}\rangle\partial_{t}\tilde{\mathbf{E}}=\nabla\times\mathbf{H}-\langle\mathbf{J}[\tilde{\mathbf{E}}]\rangle-\delta\mathbf{J}_{\perp}\label{eq:INCA-AMPERE}
\end{equation}

We note that apart from the extra current term $\delta J_{\perp}$
we now have recovered the volume-averaged curl equation for the electric
field. The correction $\delta\mathbf{J}_{\perp}=\mathbf{n}\delta J_{\perp}$
compensates the error that arises from using volume-averaged permittivities
and current densities for the normal components. Assuming an isotropic
response one obtains after some algebra 
\begin{equation}
\delta J_{\perp}=-f_{1}J_{1}[\tilde{\mathbf{E}}]-f_{2}J_{2}[\tilde{\mathbf{E}}]+\zeta_{1}J_{1}[E_{\perp,1}]+\zeta_{2}J_{2}[E_{\perp,2}]\label{eq:CURRENT-CORR-1}
\end{equation}
for the surface current correction. Its calculation requires the scalar
fields $E_{\perp,1/2}$ that are obtained by projection
\begin{equation}
\begin{aligned}E_{\perp,1} & =f_{1}\varepsilon_{\infty,1}^{-1}(\langle\varepsilon_{\infty}\rangle\mathbf{n}\cdot\tilde{\mathbf{E}}+f_{1}^{-1}\zeta_{2}\rho)\\
E_{\perp,2} & =f_{2}\varepsilon_{\infty,2}^{-1}(\langle\varepsilon_{\infty}\rangle\mathbf{n}\cdot\tilde{\mathbf{E}}-f_{2}^{-1}\zeta_{1}\rho)
\end{aligned}
\label{eq:EPERP12}
\end{equation}
Inserting these relations into (\ref{eq:CURRENT-CORR-1}) yields
\begin{equation}
\begin{split}\delta J_{\perp}= & f_{1}(J_{1}^{*}[\tilde{\mathbf{E}},\rho]-J_{1}[\tilde{\mathbf{E}}])+f_{2}(J_{2}^{*}[\tilde{\mathbf{E}},\rho]-J_{2}[\tilde{\mathbf{E}}])\end{split}
\label{eq:CURRENT-CORR-2}
\end{equation}
where we defined
\begin{equation}
J_{1/2}^{*}[\tilde{\mathbf{E}},\rho]=\zeta_{1/2}\varepsilon_{\infty,1/2}^{-1}(\langle\varepsilon_{\infty}\rangle J_{1/2}[\tilde{\mathbf{E}}]\pm f_{1/2}^{-1}\zeta_{2/1}J_{1/2}[\rho])\label{eq:J-STAR}
\end{equation}
This implies that the electric current correction can be calculated
from the currents induced by $\tilde{\mathbf{E}}$ and $\rho$. The
terms in (\ref{eq:CURRENT-CORR-2}) proportional to $J_{1/2}[\tilde{\mathbf{E}}]$
are the volume-averaged normal currents, which need to be subtracted
from eq. (\ref{eq:INCA-AMPERE}) before adding the correct $J_{1/2}^{*}[\tilde{\mathbf{E}},\rho]$
contributions. Applying (\ref{eq:EPERP12}) to the equation for the
charge field $\rho$ {[}see (\ref{eq:DEINEGA-1}){]} gives 
\begin{equation}
\begin{aligned}(f_{1}f_{2})^{-1}\partial_{t}\rho= & \zeta_{2}^{-1}J_{2}^{*}[\tilde{\mathbf{E}},\rho]-\zeta_{1}^{-1}J_{1}^{*}[\tilde{\mathbf{E}},\rho]\end{aligned}
\label{eq:SURF-CHARGE}
\end{equation}
In order to complete the update of the magnetic field the correct
electric field $\mathbf{E}$ needs to be recovered from $\tilde{\mathbf{E}}$.
This is achieved by introducing a corrective magnetic current density
\begin{equation}
\begin{split}\delta\mathbf{K}_{||} & =\nabla\times\mathbf{n}\delta E_{\perp}\\
 & =\nabla\times\mathbf{n}(\langle\varepsilon_{\infty}^{-1}\rangle\langle\varepsilon_{\infty}\rangle-1)\mathbf{n}\cdot\tilde{\mathbf{E}}
\end{split}
\label{eq:K-CURR_CORR}
\end{equation}
to Faraday's law
\begin{equation}
\partial_{t}\mathbf{H}=-\mu_{0}^{-1}\nabla\times\tilde{\mathbf{E}}-\mu_{0}^{-1}\delta\mathbf{K}_{||}\label{eq:INCA-FARADAY}
\end{equation}
This completes our reformulation of the effective cell-averaged Maxwell's
equations. The curl equations (\ref{eq:INCA-AMPERE}), (\ref{eq:INCA-FARADAY})
together with the electric and magnetic current corrections (\ref{eq:CURRENT-CORR-2})
and (\ref{eq:K-CURR_CORR}) and the surface charge equation (\ref{eq:SURF-CHARGE})
form a closed set of equations. We achieved our goal of finding an
effective medium formulation where the corrective current densities
$\delta\mathbf{J}_{\perp}$ and $\delta\mathbf{K}_{||}$ depend on
$\tilde{\mathbf{E}}$ in a functional fashion. The corrections apply
at interface cells only and vanish whenever permittivities vary smoothly
across cells. The magnetic current correction $\delta\mathbf{K}_{||}$
accounts for field discontinuities caused by a jump in the static
permittivity across the interface, while the electric current correction
$\delta\mathbf{J}_{\perp}$ captures all discontinuities induced by
the dispersive material response. Notably, calculating the induced
corrections requires only three additional physical fields, namely
the interface charge field $\rho$ and the associated induced normal
currents $J_{1/2}[\rho]$.

\section{Yee-compatible Corrective-Current FDTD scheme}

\begin{figure}
\begin{centering}
\includegraphics[scale=0.95]{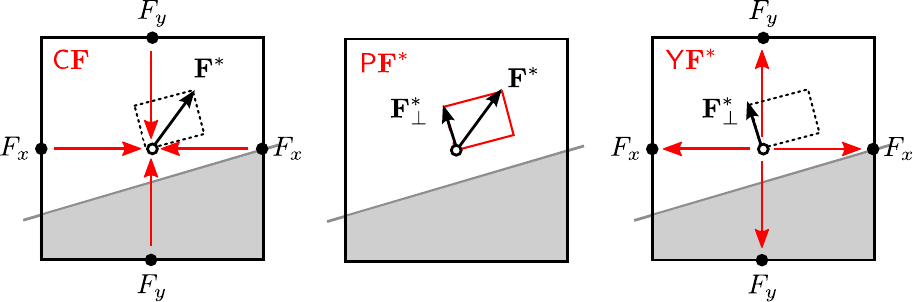}
\par\end{centering}

\caption{\label{fig:2}(color online) Pictorial representation of the action
of the $\mathsf{C}$, $\mathsf{P}$ and $\mathsf{Y}$ operators in
the two-dimensional case. $\mathsf{C}$ interpolates the vector of
Yee-centered components $\mathbf{F}$ to the cell-centered vector
$\mathbf{F}^{*}$ (left), $\mathsf{P}$ projects a cell-centered vector
onto the face normal (center), and $\mathsf{Y}$ interpolates a cell-centered
vector back onto the Yee-grid (right).}
\end{figure}

We now proceed to translate the equations derived in the previous
section into a versatile and efficient FDTD scheme. In compliancy
with the standard Yee-scheme we integrate (\ref{eq:INCA-AMPERE})
and (\ref{eq:INCA-FARADAY}) in two distinct half-steps by first performing
the electric field update
\begin{equation}
\begin{split}\tilde{\mathbf{E}}^{n+1/2}= & \tilde{\mathbf{E}}^{n-1/2}+\Delta t\langle\varepsilon_{\infty}\rangle^{-1}\nabla\times\mathbf{H}^{n}\\
 & -\Delta t\langle\varepsilon_{\infty}\rangle^{-1}(\langle\mathbf{J}^{n}[\tilde{\mathbf{E}}^{n-1/2}]\rangle+\delta\mathbf{J}_{\perp}^{n})
\end{split}
\label{eq:STEP-E}
\end{equation}
and then the magnetic field update
\begin{equation}
\begin{split}\mathbf{H}^{n+1}= & \mathbf{H}^{n}-\Delta t\mu_{0}^{-1}(\nabla\times\tilde{\mathbf{E}}^{n+1/2})+\Delta t\mu_{0}^{-1}\delta\mathbf{K}_{||}^{n+1/2}\end{split}
\label{eq:STEP-H}
\end{equation}
To keep the notation compact, we implicitly assume that $\tilde{\mathbf{E}}^{n+1/2}$
and $\mathbf{H}^{n}$ are $3N$-dimensional vectors ($N$ being the
number of Yee-cells) aggregating the electric and magnetic field components
on the staggered subgrids across the problem domain. In this formulation
the curl-operator $\nabla\times$ is a matrix that performs a stencil
operation at each point of either the electric or magnetic subgrid.
Note, that discretization turns the inverse of the volume averaged
permittivity $\langle\varepsilon_{\infty}\rangle^{-1}$ (a scalar
field) into a 3N x 3N dimensional diagonal matrix, which can be precalculated
by volume-averaging the permittivities at the various positions of
the Yee-cube. In a similar way $\langle\mathbf{J}^{n}[\tilde{\mathbf{E}}^{n-1/2}]\rangle$
can be obtained by weighting the contributing current vectors $\mathbf{J}_{i}^{n}$
with the matrix of precalculated cell-filling factors $f_{i}$. It
is important to note that the treatment of dispersive currents requires
a preceding evaluation of the response functionals $\mathbf{J}_{i}^{n}[\dots]$,
by either integrating appropriate auxiliary differential equations
(e.g., for the Lorentz pole) \cite{Taflove2005} or by using the piecewise
linear recursive convolution (PLRC) method \cite{Kelley1996}. 

Following the arguments laid out in the previous section it is clear
that the corrective currents $\delta\mathbf{J}_{\perp}^{n}$ and $\delta\mathbf{K}_{||}^{n+1/2}$
vanish whenever the material constants vary smoothly across cells.
For these volume cells $\tilde{\mathbf{E}}\rightarrow\mathbf{E}$
and the update equations reduce themselves to the dispersive VA FDTD
method, which, as $\langle\varepsilon_{\infty}\rangle^{-1}$ is diagonal
can be efficiently integrated using the standard Yee-scheme. Within
interface cells, on the other hand, $\tilde{\mathbf{E}}$ differs
from the electric field $\mathbf{E}$ and a corrective step is necessary
to accurately account for the discontinuity of the normal field component.
As shown before the discontinuity in the normal component is directly
proportional to the surface charge density $\rho$ induced at the
interface. Discretizing (\ref{eq:SURF-CHARGE}) results in an update
equation for $\rho$
\begin{equation}
\begin{aligned}\rho^{n+1/2}=\rho^{n-1/2} & +\Delta t(f_{1}f_{2})(\zeta_{2}^{-1}J_{2}^{*n}-\zeta_{1}^{-1}J_{1}^{*n})\end{aligned}
\label{eq:SURF-CHARGE-1}
\end{equation}

that requires evaluation of the currents $J_{1/2}^{*n}$ according
to (\ref{eq:J-STAR}). In difference to the electromagnetic field
components, which are evaluated on the Yee-grid, $\rho$ is a cell-centered
quantity. We therefore need to introduce operators to interpolate
between the Yee- and cell-centered grids. Figure \ref{fig:2} illustrates
the action of the $\mathsf{Y}$ and $\mathsf{C}$ interpolation operators
(left and right panel) together with the projection operator $\mathsf{P}$
(center panel). Applied to write (\ref{eq:J-STAR}) this yields
\begin{equation}
\begin{split}J_{1/2}^{*n}=\zeta_{1/2}\varepsilon_{\infty,1/2}^{-1} & (\mathsf{P}\mathsf{C}\langle\varepsilon_{\infty}\rangle\mathbf{J}{}_{1/2}^{n}[\tilde{\mathbf{E}}^{n-1/2}]\\
 & \pm f_{1/2}^{-1}\zeta_{2/1}J_{1/2}^{n}[\rho^{n-1/2}])
\end{split}
\label{eq:J-STAR-2}
\end{equation}

This expression recycles the previously calculated $\mathbf{J}{}_{1/2}^{n}$
currents on the Yee-grid but introduces a charge-current $J_{1/2}^{n}[\rho^{n-1/2}]$
that, using the same current-functional, is evaluated at the cell
center. To improve smoothness of the fields under the projection/interpolation
operation we multiply $\mathbf{J}{}_{1/2}^{n}$ with the $\langle\varepsilon_{\infty}\rangle$
tensor, which is already available on the Yee-grid. In contrast, the
coefficients $\zeta_{1/2}$, $\varepsilon_{\infty,1/2}^{-1}$ and
$f_{1/2}^{-1}$ and the face-normal $\mathbf{n}$ are parameters that
are defined at the cell-center (see Fig. \ref{fig:computational_cost}).
As (\ref{eq:J-STAR-2}) can be evaluated on-the-fly, the only additional
physical fields that need to be stored at the cell-center are $\rho$
and its induced currents $J_{1/2}^{n}[\rho^{n-1/2}]$.

With the surface charge and its currents known, it becomes possible
to compute the corrections $\delta\mathbf{J}_{\perp}^{n}$ and $\delta\mathbf{K}_{||}^{n+1/2}$
that enter the update equations (\ref{eq:STEP-E}) and (\ref{eq:STEP-H}).
However, the order of operators (and hence the discretization) is
ambiguous, and, as the scheme is corrective, can impact on the stability
of the scheme. A numerical analysis of the computational errors suggests
that $\mathbf{J}{}_{1/2}^{n}$ is best multiplied with the $\langle\varepsilon_{\infty}\rangle$
tensor before centering to the grid. This is due to the fact that
the normal component of $\langle\varepsilon_{\infty}\rangle\tilde{\mathbf{E}}$
retains smoothness across adjacent cells with different $\varepsilon_{\infty}$.
Further, to maintain consistency between the Yee and cell-centered
update equations (\ref{eq:STEP-E}), (\ref{eq:STEP-H}) and (\ref{eq:SURF-CHARGE-1})
we assign parameters as indicated by Fig. \ref{fig:computational_cost}.
This allows us to write
\begin{equation}
\begin{split}\langle\varepsilon_{\infty}\rangle^{-1}\delta\mathbf{J}_{\perp}^{n}= & -\langle\varepsilon_{\infty}\rangle^{-2}(f_{1}\mathsf{Y}\mathsf{P}\mathsf{C}\langle\varepsilon_{\infty}\rangle\mathbf{J}_{1}^{n}-f_{2}\mathsf{Y}\mathsf{P}\mathsf{C}\langle\varepsilon_{\infty}\rangle)\mathbf{J}_{2}^{n}\\
 & +\mathsf{Y}\mathbf{n}f_{1}J_{1}^{*n}+\mathsf{Y}\mathbf{n}f_{2}J_{1}^{*n}
\end{split}
\label{eq:CURRENT-CORR-2-1}
\end{equation}
where volume filling factors in the first line are applied after centering
onto the Yee-grid, and, for the second line, directly at the cell-center. 

The discretization of the magnetic current requires both terms in
(\ref{eq:K-CURR_CORR}) to be interpolated to the center before spreading
them out again onto the Yee-grid. We obtain
\begin{eqnarray}
\delta\mathbf{K}_{||}^{n+1/2} & = & \nabla\times\delta\tilde{\mathbf{E}}_{\perp}^{n+1/2}\label{eq:K_CORR-2}
\end{eqnarray}
with
\begin{equation}
\delta\tilde{\mathbf{E}}_{\perp}^{n+1/2}=(\mathsf{Y}\langle\varepsilon_{\infty}^{-1}\rangle-\langle\varepsilon_{\infty}\rangle^{-1}\mathsf{Y})\mathsf{P}\mathsf{C}\langle\varepsilon_{\infty}\rangle\tilde{\mathbf{E}}^{n+1/2}
\end{equation}

Fundamentally, both the electric and magnetic current corrections
can be calculated on-the-fly. As the corrections only apply to interface
cells, they can be added in a separate step to the update equations.
This means that the update equations of the VA FDTD scheme can be
deployed across the whole grid, followed by oversampling steps that
perform the current-corrections (CC) for interface cells only. The
complete update sequence for the VA+CC algorithm is shown in Alg.
\ref{alg:DVA+CC}.

\begin{algorithm}
\begin{itemize}
\item \emph{n+1/2 (on Yee-grid):}

\begin{itemize}
\item VA: update $\tilde{\mathbf{E}}^{n-1/2}\rightarrow\tilde{\mathbf{E}}^{n+1/2}$
w/o $\delta\mathbf{J}_{\perp}^{n}$ {[}(\ref{eq:STEP-E}){]}
\item CC: add correction $\delta\mathbf{J}_{\perp}^{n}$ {[}(\ref{eq:CURRENT-CORR-2-1}){]}
\end{itemize}
\item \emph{n+1/2 (on centered-grid):}

\begin{itemize}
\item CC: update $\rho^{n-1/2}\rightarrow\rho^{n+1/2}$ {[}(\ref{eq:SURF-CHARGE-1}){]}
\item CC: evaluate $J_{i}^{n+1}[\rho^{n+1/2}]$ (for next cycle)
\end{itemize}
\item \emph{n+1 (on Yee-grid)}

\begin{itemize}
\item VA: evaluate $\mathbf{J}_{i}^{n+1}[\tilde{\mathbf{E}}^{n+1/2}]$ (for
next cycle)
\item VA: update $\mathbf{H}^{n}\rightarrow\mathbf{H}^{n+1}$ w/o $\delta\mathbf{K}_{||}^{n+1/2}$
{[}(\ref{eq:STEP-H}){]}
\item CC: add correction $\delta\mathbf{K}_{||}^{n+1/2}$ {[}(\ref{eq:K_CORR-2}){]}
\end{itemize}
\end{itemize}
\caption{\label{alg:DVA+CC}Sequence of field updates (VA+CC)}
\end{algorithm}

As each step can be associated with a loop over cells, it becomes
evident that the current-correction (CC) steps augment those related
to the VA FDTD scheme. As the CC steps only apply to interface cells,
the computational overhead of the VA+CC FDTD scheme is not significant
unless the number of interface cells becomes comparable to the number
of volume cells.

\section{Results}

\begin{figure}
\begin{centering}
\includegraphics[width=0.95\linewidth]{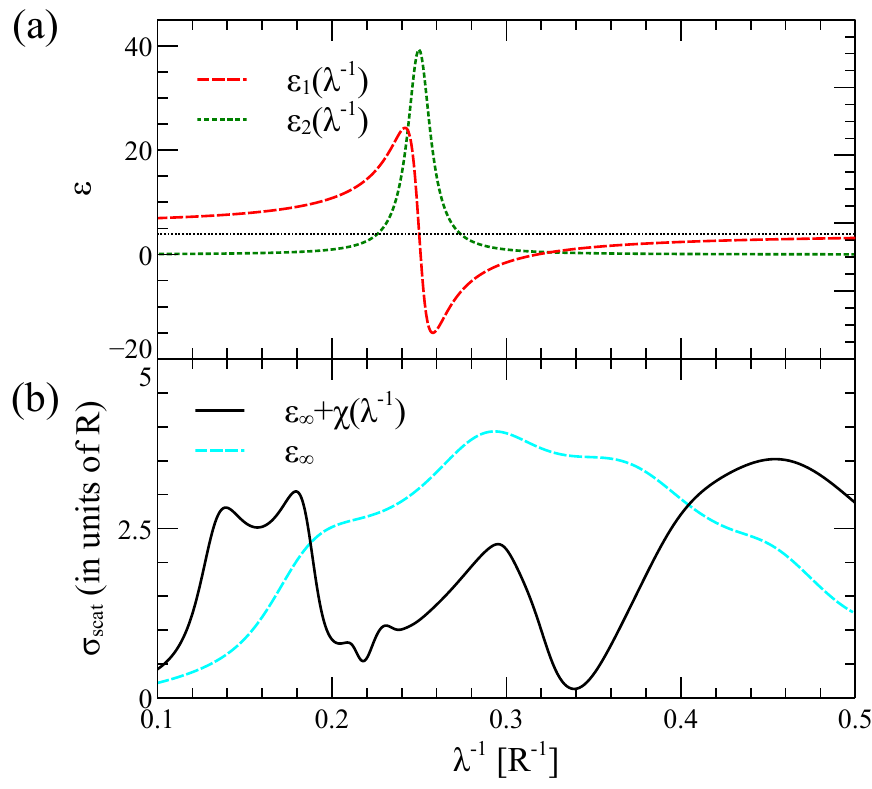}
\par\end{centering}

\caption{\label{fig:dielectric_function}(color online) (a) the real (dashed)
and imaginary (dotted) part of the dielectric function of an infinitely
long cylinder and (b) the analytically calculated scattering cross
section (solid). The dotted curve in (b) represents the scattering
cross section of a cylinder with a purely static dielectric constant
of $\varepsilon_{\infty}=4$ (indicated by the thin dotted line in
(a)}
\end{figure}

To verify the accuracy of our method we compare our numerical calculations
with the Mie scattering cross section of an infinitely extended strongly
dispersive cylinder excited by a TM plane-wave. Although the calculations
presented here are 2D, the derived equations and algorithms are also
valid in 3D. The dielectric function describing the response of the
cylinder consists of a single Lorentzian resonance at $\lambda_{0}^{-1}=0.25R$
and a background dielectric constant of $\varepsilon_{\infty}=4$,
where $R$ is the radius of the cylinder. Figure \ref{fig:dielectric_function}a
shows real and imaginary parts of the complex permittivity $\varepsilon(\lambda^{-1})=\varepsilon'(\lambda^{-1})+i\varepsilon''(\lambda^{-1})=\varepsilon_{\infty}+2.5\lambda_{0}^{-2}(\lambda_{0}^{-2}-\lambda^{-2}-i0.05\pi^{-1}\lambda^{-1})^{-1}$
together with the analytically calculated scattering cross-sections
for scatterers with and without the dispersive contribution $\chi(\lambda^{-1})$
(Fig. \ref{fig:dielectric_function}b).

\begin{figure}
\begin{centering}
\includegraphics[width=0.5\linewidth]{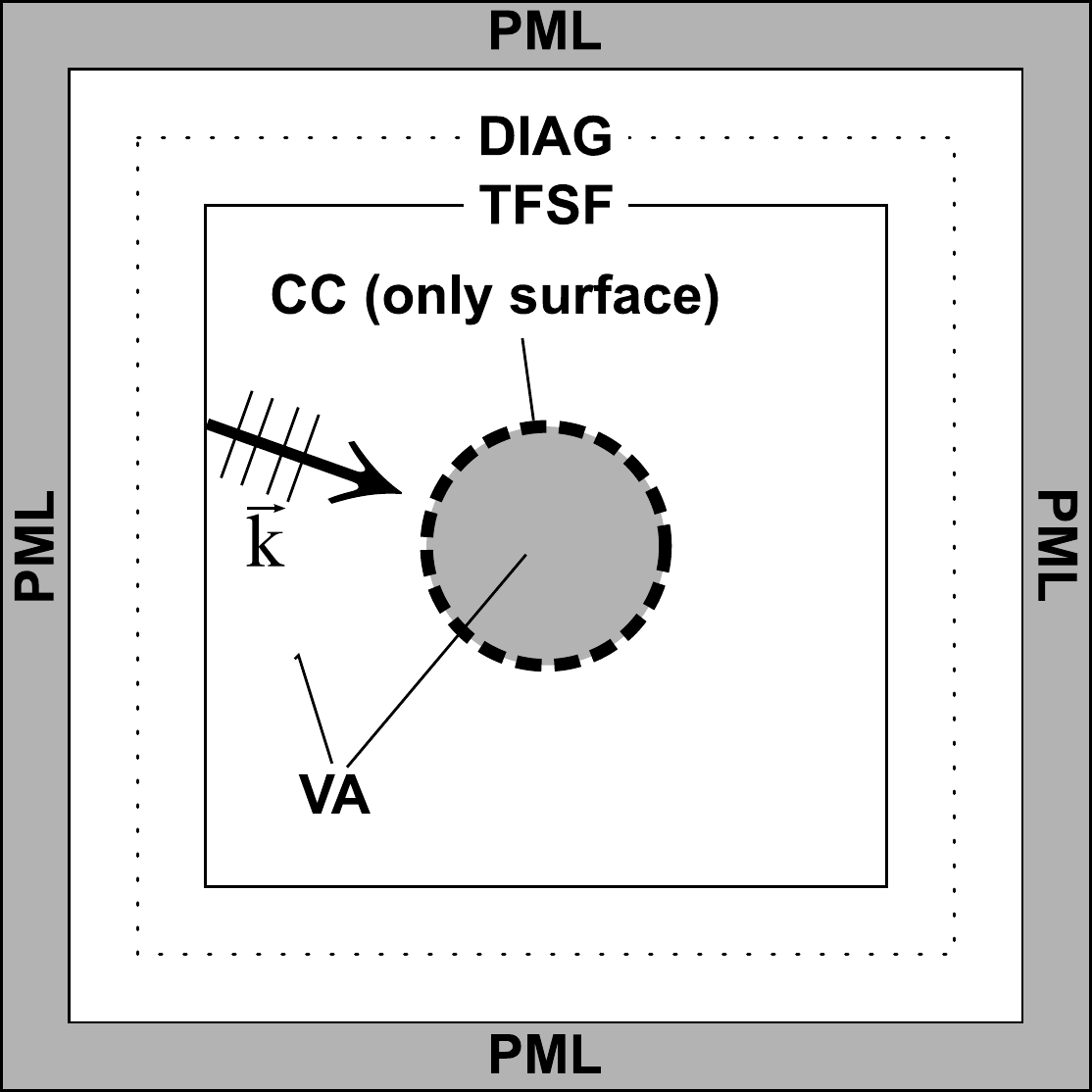}
\par\end{centering}

\caption{\label{fig:numerical_setup}Computational setup: an incident pulse
is injected on the inner left boundary of a TFSF box. The pulse interacts
with the scatterer and leaves the TFSF box outside of which the energy
flux of the scattered field is recorded (dotted line marked ``DIAG'').
The boundary of the computational region is terminated with perfectly
matched layers (PML). The current corrections (CC) $\delta\mathbf{J}_{\perp}$
and $\delta\mathbf{K}_{||}$ are only applied at the surface of the
cylinder (thick dashed line).}
\end{figure}

\begin{figure}
\begin{centering}
\includegraphics[width=0.95\linewidth]{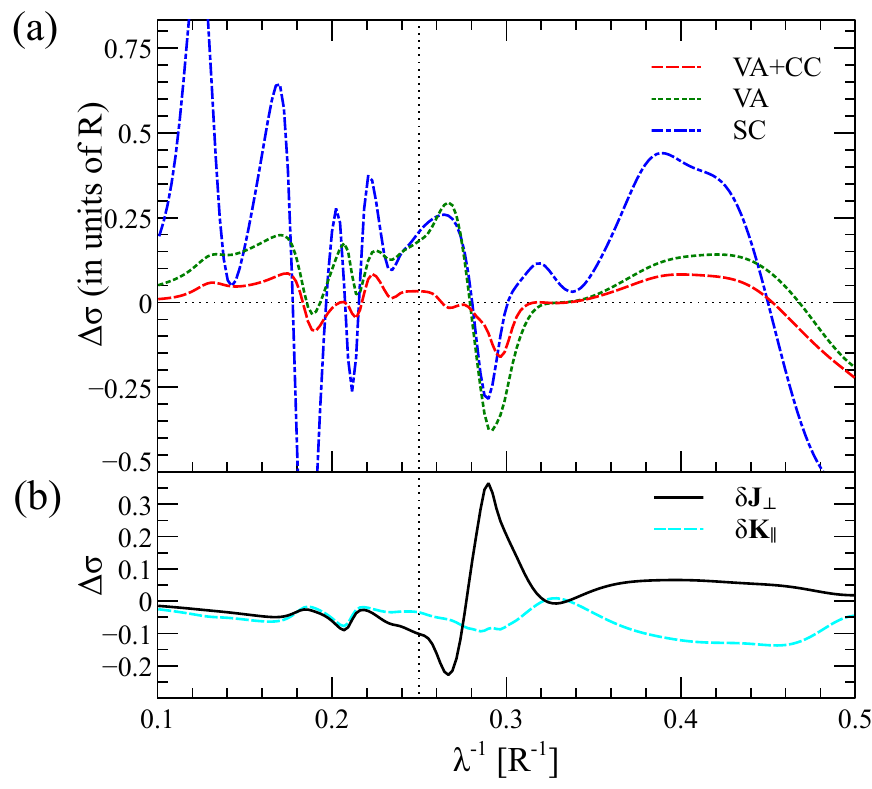}
\par\end{centering}

\caption{\label{fig:numerical_spectrum}(color online) (a) difference between
the analytic and numerical scattering cross section spectrum of an
infinitely long dispersive Mie cylinder calculated with a VA FDTD
scheme with (red dashed) and without (green dotted) charge corrections.
The result of a staircased FDTD scheme is shown for reference (dashed-dotted
blue). The thin dotted black horizontal line is a guide to the eye.
(b) the contribution of the corrective electric and magnetic currents
to the cross section spectrum}
\end{figure}

The numerical setup of the 2D calculation is depicted in Fig. \ref{fig:numerical_setup}.
A Total-Field-Scattered-Field (TFSF) box \cite{Taflove2005} is used
to inject pulses with $\mathbf{E}_{inc}(\mathbf{r},t)=\mathbf{E}_{0}A(t)\exp(-i\omega t+i\mathbf{k}\cdot\mathbf{r})$
with temporal envelope $A(t)$, polarisation $\mathbf{E}_{0}$ and
center frequency $\omega=c|\mathbf{k}|$ into the system in direction
of $\mathbf{k}$ (where $\mathbf{k}\perp\mathbf{E}_{0}$). To minimize
the error from numerical dispersion we take into account the numerical
phase velocity at the center frequency for the given angle of incidence
and chose a sufficiently narrow-band excitation. The energy flux $\mathbf{E}\times\mathbf{H}$
of the scattered field is recorded at the boundary of a box located
outside of the TFSF box. The computational region is terminated with
perfectly matched layers (PML) \cite{Taflove2005} which nearly completely
attenuate any reflections caused by the computational boundary. After
the simulation, the scattering spectrum can be retrieved by Fourier-transforming
the fields recorded at a closed surface outside of the TFSF box (marked
with DIAG in Fig. \ref{fig:numerical_spectrum}).

\begin{figure}
\begin{centering}
\includegraphics[width=0.95\linewidth]{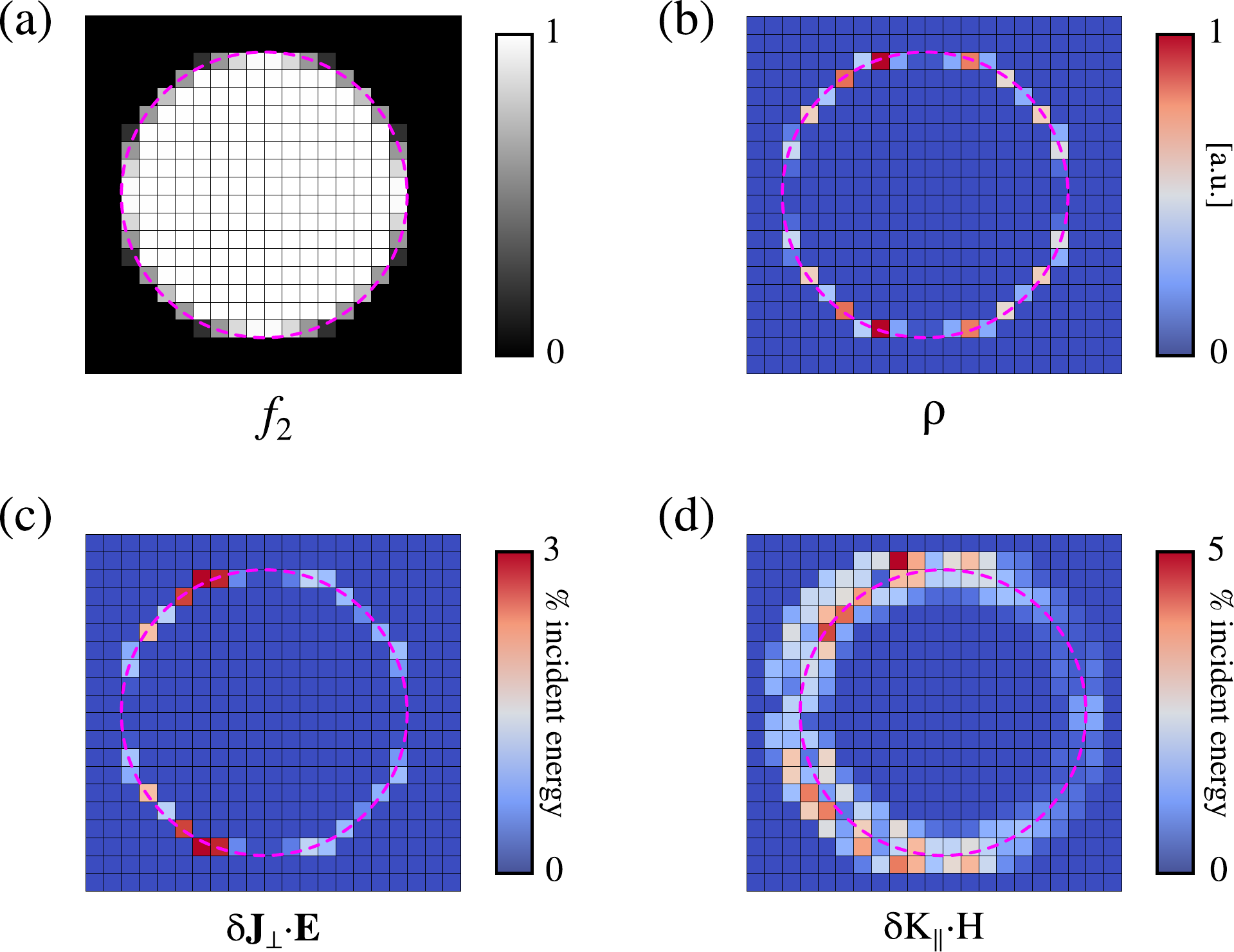}
\par\end{centering}

\caption{\label{fig:field_plot}(color online) Snapshot of (a) the volume filling
factor $f_{2}$ of the interface cells (b) the charge field $\rho$
(c) the electric current correction energy $\delta\mathbf{J}_{\bot}\cdot\mathbf{E}$
(d) the magnetic current correction energy $\delta\mathbf{K}_{\parallel}\cdot\mathbf{H}$.}
\end{figure}

Figure \ref{fig:numerical_spectrum} (top) shows the difference between
the analytic and numerical scattering cross sections obtained by numerical
simulation with a resolution of eight Yee-cells per cylinder radius.
The results of the VA+CC FDTD scheme (dashed red line) are in better
agreement with the analytical calculation than the VA FDTD scheme
(dotted green line) throughout the spectrum. For comparison, the result
of a simple staircased FDTD scheme was included in the figure (dash-dotted
blue line). By selectively disabling either the current correction
$\delta\mathbf{J}_{\perp}$ or $\delta\mathbf{K}_{||}$ and subtracting
the result from the VA FDTD scheme, the contributions of the charge
corrections to the spectrum were quantified (Fig. \ref{fig:numerical_spectrum}b).
The contribution $\delta\mathbf{J}_{\perp}$ shows a prominent peak
at a frequency which is slightly offset to the resonance frequency
of the Lorentzian (indicated by the vertical dotted line). To illustrate
the spatial dependence of the corrections and the charge density we
plot contour images of the charge field $\rho$ (Fig. \ref{fig:field_plot}b),
the energy density of the electric correction $\delta\mathbf{J}_{\bot}\cdot\mathbf{E}$
(Fig. \ref{fig:field_plot}c), and the energy density of the magnetic
correction $\delta\mathbf{K}_{\parallel}\cdot\mathbf{H}$ (Fig. \ref{fig:field_plot}d).
Whereas the corrections associated with the charge density and electric
field correction are stored at the cell center, the correction associated
with $\delta\mathbf{K}_{\parallel}\cdot\mathbf{H}$ is calculated
from Yee-centered quantities and therefore appears to be smeared out
over several adjacent cells.

\begin{figure}
\begin{centering}
\includegraphics[width=0.9\linewidth]{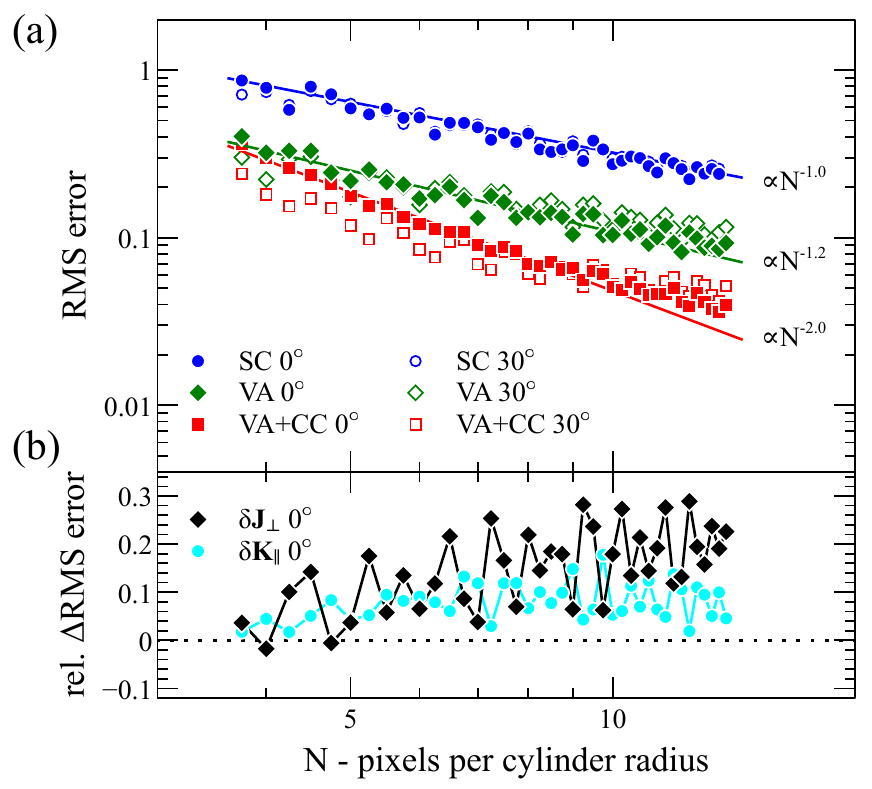}
\par\end{centering}

\caption{\label{fig:error_plot}(color online) RMS error vs. increasing pixels
per cylinder radius $N$. (a) the RMS error of staircasing (blue circles),
VA (green diamonds) and VA+CC FDTD (red squares) schemes for incident
angles $0^{\circ}$ (filled) and $30^{\circ}$ (hollow) (b) the relative
change in error when enabling either the $\delta\mathbf{J}_{\perp}$
(cyan circles) or $\delta\mathbf{K}_{||}$ (black diamonds) current
corrections compared to the VA FDTD scheme error.}
\end{figure}

To investigate the convergence behavior of the charge correction algorithm,
numerical simulations with increasing resolution $N$ were conducted
for incident angles of $0^{\circ}$ and $30^{\circ}$. The RMS error
for each simulation was obtained, by comparing the numerical scattering
cross section spectrum with the analytical result (Fig. \ref{fig:error_plot}a).
The overall error reduction is achieved by the combined action of
the corrections $\delta\mathbf{J}_{\perp}$ and $\delta\mathbf{K}_{||}$
as shown in Fig. \ref{fig:error_plot}b. The VA (green diamonds) and
the staircasing (blue circles) FDTD scheme produce errors that are
significantly larger than those of the VA+CC scheme (red squares),
whose RMS error decreases with $\propto N^{-2.0}$. For higher resolutions
the decrease in error saturates, which may be attributed to error
contributions from the PMLs. We therefore conclude that for this particular
system VA+CC is approximately second order accurate and consistently
achieves lower errors than the VA scheme.

Finally, we compare the computational cost (memory and processing
time) for the different schemes. The results are summarized in Fig.
\ref{fig:computational_cost}. The staircasing scheme only requires
the static epsilon $\varepsilon_{\infty}$ at each Yee-cell position
of the $\mathbf{E}$-field and the three vectorial fields $\mathbf{E}$,
$\mathbf{H}$, $\mathbf{J}_{1}$. The VA algorithm additionally stores
the filling factors $f_{1}$ at each Yee-cell position of the $\mathbf{E}$-field.
The VA+CC scheme is identical to the VA scheme for non-interface cells
requiring $15$ scalar components. At interface cells the VA+CC scheme
requires an additional $7$ scalar components for storing $\rho$,
$J_{1}[\rho]$, $\mathbf{n}$, $f_{1}$ and $\varepsilon_{\infty,1/2}$.
The comparision of computation time indicates an almost identical
performance for the staircase and VA schemes. VA+CC delivers the same
performance for volume cells but requires additional computational
steps for interface cells, resulting in $\approx50\%$ overhead in
the per cell processing time. These overheads seems significant but
rarely matter for practical applications as the surface to volume
ratio is typically small. For the Mie scattering simulations presented
in Fig. \ref{fig:numerical_spectrum} for example (8 cells per radius)
the increase in computation time of the VA+CC algorithm is $<1\%$
(compared to VA) as the interface/volume cell ratio is $\approx0.6\%$.

\begin{figure}
\begin{centering}
\begin{tabular}{|l|c|c|c|}
\hline 
 & Staircase & VA & VA+CC\tabularnewline
\hline 
\hline 
\multirow{2}{*}{fields (YG)} & $\mathbf{E}$, $\mathbf{H}$, $\mathbf{J}_{1}$ & $\mathbf{E}$, $\mathbf{H}$, $\mathbf{J}_{1}$ & $\tilde{\mathbf{E}}$, $\mathbf{H}$, $\mathbf{J}_{1}$\tabularnewline
\cline{2-4} 
 & $\varepsilon_{\infty}$ & $\langle\varepsilon_{\infty}\rangle$, $f_{1}$ & $\langle\varepsilon_{\infty}\rangle$, $f_{1}$\tabularnewline
\hline 
\multirow{2}{*}{fields (CG)} & - & - & $\rho$, $J_{1}[\rho]$\tabularnewline
\cline{2-4} 
 & - & - & $\mathbf{n}$, $f_{1}$, $\varepsilon_{\infty,1/2}$\tabularnewline
\hline 
storage/cell & 12 & 15 & 15+7\tabularnewline
\hline 
time/cell & 1.67 & 1.67 & 1.67+0.83\tabularnewline
\hline 
\end{tabular}
\par\end{centering}

\caption{\label{fig:computational_cost}Computational cost (memory and CPU)
of staircasing, VA and VA+CC FDTD algorithms. The various fields and
parameters are either assigned to cell-centered (CG) positions or
to the Yee-grid (YG). VA+CC requires same storage as VA for volume
cells but carries an overhead of \textasciitilde{}50\% for interface
cells. The per-cell storage values are given in QWORDs, the per-cell
time in microseconds. }

\end{figure}

\section{Conclusion}

In summary we presented an effective-medium theory that takes a corrective
approach to the cell-averaged Maxwell's curl equations. The theory
holds for static and linear dispersive permittivities and captures
the field discontinuities inside a cell in form of surface current
corrections, which can be calculated by integrating a surface charge
equation alongside the volume-averaged curl equations. We derived
a computationally efficient FDTD algorithm that allows deploying the
standard Yee-algorithm across the domain followed by surface current
corrections that selectively apply at interface cells. The improvement
in accuracy is quantified by calculating spectral scattering cross-sections
of strongly dispersive Mie scatterers. The extracted error exponents
indicate that the algorithm approximately restores second order accuracy.
The work presented is relevant in the current context of nano-photonic
research and may pave the way to the development of novel pertubative
techniques for solving Maxwell's equations.

We acknowledge useful discussions with Andrew Horsefield. This work
was supported by the Leverhulme Trust and the UK Engineering and Physical
Sciences Research Council.

\bibliographystyle{IEEEtran}

\begin{IEEEbiography}[{\includegraphics[width=1in]{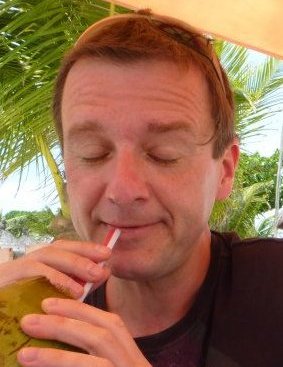}}]{Dr. Joachim Hamm}
is a Leverhulme research fellow for Plasmonics and Metamaterials at Imperial College London. He did his PhD at the German AerospaceCentre(DLR), where he designed and implemented parallel software on supercomputers targeting the efficient microscopic modelling of vertical cavity surface emitting lasers (VCSELs). His research interests focus on the investigation of functional (active and nonlinear) metamaterial design, extreme light-matter interaction on the nanoscale and the stopping and localisation of light in solid-state structures.
\end{IEEEbiography}

\begin{IEEEbiography}[{\includegraphics[width=1in]{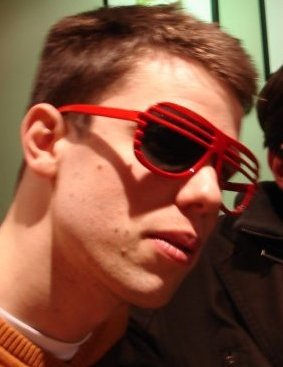}}]{Fabian Renn} received his diploma degree from the University of Heidelberg, Germany in 2010, followed by an Msc degree from Imperial College London in 2011. Currently, he is a PhD student within the group of Prof. O. Hess at Imperial College London. His research interests include numerical simulation techniques of light matter interactions. His diploma thesis focused on theory and simulation of x-rays interacting with patterned strained silicon germanium at AMD Dresden, Germany.
\end{IEEEbiography}

\begin{IEEEbiography}[{\includegraphics[width=1in]{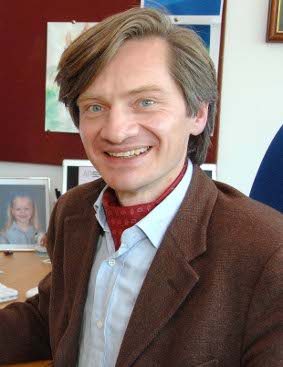}}]{Prof. Ortwin Hess} holds the Leverhulme Chair in Metamaterials in the Department of Physics at Imperial College London and is Co-Director of the Centre for Plasmonics \& Metamaterials. Ortwin studied physics at the University of Erlangen and the Technical University of Berlin. Ortwin has been (from 1995 to 2003) Head of the Theoretical Quantum Electronics Group at the Institute of Technical Physics in Stuttgart, Germany. Since 2001 he is Docent of Photonics at Tampere University of Technology in Finland. Ortwin has been Visiting Professor at Stanford University (1997 - 1998) and the University of Munich (2000 - 2001). From 2003-2010 he held the Chair of Theoretical Condensed Matter and Optical Physics in the Department of Physics and the Advanced Technology Institute at the University of Surrey in Guildford, UK where he is now a Visiting Professor. 
\end{IEEEbiography}
\vfill
\end{document}